\documentstyle[prl,aps,epsf]{revtex}

\begin{document}

\twocolumn[\hsize\textwidth\columnwidth\hsize\csname @twocolumnfalse\endcsname

\title{Velocity Selection for Propagating Fronts in Superconductors}

\author{S.\ John Di Bartolo\cite{sjdemail} and Alan T.\ Dorsey\cite{atdemail}}

\address{Department of Physics,  University of Virginia, McCormick Road,
Charlottesville, Virginia 22901}

\date{\today}

\maketitle

\begin{abstract}

Using the time-dependent Ginzburg-Landau equations we study the 
propagation of planar fronts in superconductors, which
would appear after a quench to zero applied magnetic field. 
Our numerical solutions show that the fronts propagate at a unique speed 
which is controlled by the amount of magnetic flux trapped in the front.
For small flux the speed can be determined from the linear
marginal stability hypothesis, while for large flux the
speed may be calculated using matched asymptotic expansions.
At a special point the order parameter and vector potential are
{\it dual},  leading to an {\it exact} solution which 
is used as the starting point for a perturbative analysis.

\end{abstract}


\vskip2pc]

There exists a wide class of problems in which a system, subjected
to a sudden destabilizing change, responds
by forming fronts which propagate into the unstable state.
This phenomena occurs in
models of population dynamics \cite{fisher37,kolmogorov,aronson78},
pulse propagation \cite{scott}, liquid crystals \cite{vansaarloos95},
solidification \cite{solidification}, and tubular vesicles \cite{goldstein96};
further examples are discussed in Ref.~\cite{cross93}.   The
interesting issues common to all of these problems are whether
the fronts approach a constant, unique speed, and if so, how the
system selects this speed out of a manifold of possible speeds.
The prototype is {\it Fisher's equation} \cite{fisher37,kolmogorov},
\begin{equation}
\partial_t u  = \partial^2_x u   + F(u),
\label{fisher}
\end{equation}
where  $u>0$ may be interpreted as a population density, and
$F(0) = F(1) = 0$.  As shown rigorously
by Aronson and Weinberger \cite{aronson78},
for a sufficiently localized initial condition the solutions of this
equation evolve into fronts of the form $u(x,t) = U(x-ct)$ which
connect $u=1$ at $x=-\infty$ to $u=0$ at $x=\infty$; the speed of the
front satisfies $2\sqrt{F'(0)}\leq c \leq 2\ {\rm sup} \sqrt{F(u)/u}$,
so that for the special case $F(u) = u - u^3$ the selected speed is
$c=2$.  There have been many attempts to generalize these results to
more complicated equations and physical systems, including heuristic
methods such as the
{\it marginal stability hypothesis} (MSH) \cite{benjacob85,vansaarloos89}
and the {\it structural stability hypothesis} \cite{paquette94,chen94},
construction of exact solutions \cite{vansaarloos89,benguria94b},
variational methods \cite{chen94,benguria94a,benguria96}, and
dynamical systems methods \cite{goriely95}.

In this paper we will study a closely related problem of front
(or interface) propagation in superconductors.  The problem which
we have in mind is the following: begin with a bulk superconductor
in an applied magnetic field equal to the critical field $H_c$, so that
there is a stationary, planar superconducting-normal interface which separates
the normal and superconducting phases; then
rapidly reduce the applied field to {\it zero}, so that the
now unstable interface propagates toward the normal phase so as to expel
any trapped magnetic flux, leaving the sample in the Meissner phase.  
Assuming that the interface remains planar,
what is its dynamics?  In this paper we will show that under these
conditions constant velocity fronts do propagate, at a unique speed
which is controlled by the amount of magnetic flux which is trapped in the 
front \cite{frahm91}.   We calculate this speed in different parameter regimes using the MSH, 
a perturbative calculation about an exact solution which we have discovered, 
and an asymptotic analysis valid for small speeds.  Where appropriate, 
our analytic results are compared to extensive numerical solutions of the 
dynamic equations. 

To analyze the behavior of the superconducting-normal boundary, we use the
one-dimensional time-dependent Ginzburg-Landau (TDGL) equations, which in 
dimensionless units \cite{dorsey94} are 
\begin{equation}
 \partial_t f = {1\over \kappa^{2}}
 \partial^2_x  f  - q^{2} f + f - f^{3},
\label{f}
\end{equation}
\begin{equation}
\bar{\sigma} \partial_t q  =  \partial^2_x q -f^{2}  q, 
\label{q}
\end{equation}
where $f$ is the magnitude of the superconducting order parameter,
$q$ is the gauge-invariant vector potential (such that $h= \partial_x q $
is the magnetic field), $\kappa$ is the
Ginzburg-Landau parameter, and $\bar{\sigma}$ is the dimensionless normal state
conductivity (the ratio of the order parameter diffusion constant to the 
magnetic field diffusion constant).   
Notice that if the vector potential is zero, then
Eq.~(\ref{f}) is exactly Fisher's equation, Eq.~(\ref{fisher}), with 
fronts which propagate at a speed $c=2/\kappa$ in our units. 
Since lengths are measured in units of the 
penetration depth (typically of order 500 \AA\ in type-I superconductors)
and time in units of the order parameter relaxation time
(of order $10^{-9}$--$10^{-10}$ s), the characteristic 
scale for speeds is of order 100 m/s \cite{leiderer93}.  
For the propagating solutions which we are
considering (i.e., after the field quench), $f = 1$, $q=0$, and 
$h=0$ as $x\rightarrow -\infty$
(the superconducting phase), and $f=0$, $q=Q_\infty$, and $h=0$
as $x\rightarrow \infty$ (the normal phase).  The physical
meaning of $Q_\infty$ is clear once we notice that the
integrated magnetic field in the
front (i.e., the total magnetic flux per unit length parallel to the front)
is $\int_{-\infty}^{\infty} h(x)\, dx = Q_\infty$.
As we will see, $Q_\infty$ is an important control parameter for
the front dynamics---the larger the trapped magnetic flux in the front
the smaller the front speed.

\begin{figure}
\centerline{
\epsfxsize=8cm \leavevmode \epsfbox{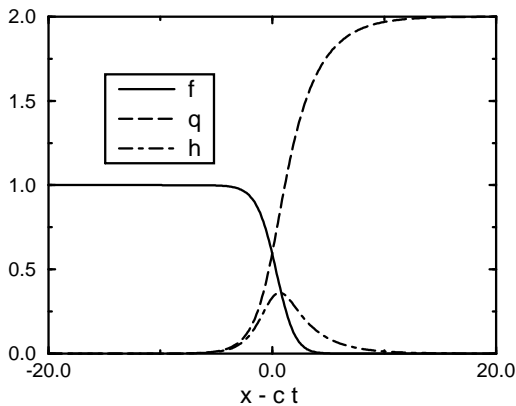}}
\caption{Numerical solution for $\kappa=1$, $\bar{\sigma}=1$,
and $Q_\infty=2$, resulting in a front which moves to the 
right with speed $c=0.386$.
Shown are the order parameter (solid line), vector potential (dashed line),
and field (dashed-dotted line).} 
\label{fig1}
\end{figure}

\begin{figure}
\centerline{
\epsfxsize=8cm \leavevmode \epsfbox{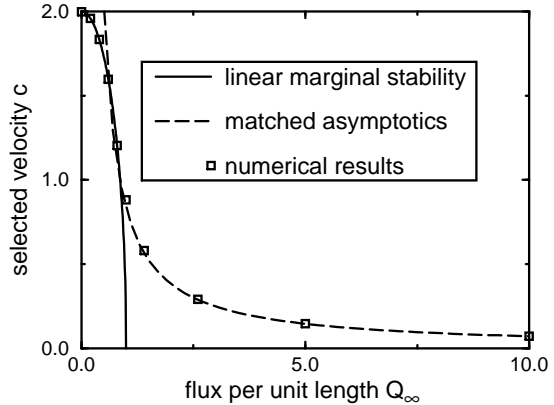}}
\caption{Numerical values (open squares) of the front speed as a
function of $Q_\infty$ for $\kappa = 1$ and $\bar{\sigma} = 1$.  
Also shown are the 
results of the linear marginal stability analysis (solid line) 
and matched asymptotic analysis (dashed line).}
\label{fig2}
\end{figure}

We have solved Eqs.~(\ref{f}) and (\ref{q}) numerically for
a wide range of $\kappa$, $\bar{\sigma}$ and $Q_\infty$, using  
the Crank-Nicholson method \cite{recipes}. 
An initial configuration for the order parameter and magnetic field
with the appropriate boundary conditions is established and then allowed to
evolve in time.  For our boundary conditions (in
particular, $h=0$ as $x\rightarrow\infty$), the front rapidly
approaches a constant velocity.
On an IBM RS 6000/370 approximately 5000 cpu minutes are needed to trace
this evolution (allowing 300-500 time units to
elapse usually brings us sufficiently close to a constant velocity
solution).
We analyze the profile of the order parameter and magnetic field for the
constant velocity solutions (see Fig.~\ref{fig1} for a representative
result) and determine the value of the front velocity as a function of
$Q_\infty$, as shown in Fig.~\ref{fig2}.  
The remainder of this paper is devoted to an analysis of the TDGL
equations which will shed some light on these numerical results.

To begin our analysis we will search for steady traveling
wave solutions of the TDGL equations,
of the form  $f(x,t) = F(X)$ and $q(x,t) = Q(X)$
where $X=x-ct$ with $c>0$. Then the TDGL equations become
\begin{equation}
{1\over \kappa^{2}} F'' + c F' - Q^2 F + F - F^3 = 0,
\label{F}
\end{equation}
\begin{equation}
Q'' + \bar{\sigma} c Q' - F^2 Q = 0,
\label{Q}
\end{equation}
(the primes denote differentiation with respect to $X$).
The order parameter connects $F=1$ at $X=-\infty$ to
$F=0$ at $X=\infty$, while the
vector potential connects $Q=0$ at $X=-\infty$ to $Q=Q_\infty$
at $X=\infty$.  In the spirit
of the MSH, we first examine the linear stability
of the leading edge of the front ($X\rightarrow \infty$).
Linearizing Eqs.~(\ref{F})
\noindent and (\ref{Q}), we find $F(X)\sim \exp(-\lambda_{-} X)$, with
\begin{equation}
\lambda_{-} =  {c\kappa^2 \over 2}- 
  \sqrt{ \left( {c\kappa^2\over 2} \right)^2 - \kappa^2 [1-Q_\infty^2]}.
\label{lambda}
\end{equation}
Since the magnitude of the order parameter $F\ge 0$, we require
${\rm Im}(\lambda_{-})=0$
in order to prevent the 
solution from oscillating, which implies 
\begin{equation}
c\geq c'=  {2 \sqrt{1 - Q_\infty^2} \over \kappa}\qquad (Q_\infty\leq 1),
\label{msh}
\end{equation}
with $c\geq 0 $ for $Q_\infty\geq 1$.   In addition to this
lower bound on the speed, we conjecture that $c\leq 2/\kappa$, 
since we expect the front to have the greatest speed in the absence
of any trapped flux.  The MSH
\cite{benjacob85,vansaarloos89} would have us set
$c=c'$; at this speed the leading edge of the order parameter front
is marginally stable with respect to perturbations---the phase and
group velocities of the perturbation are equal at this speed.
At $Q_\infty=0$, this yields $c = 2/\kappa$, which is the
rigorous result for Fisher's equation \cite{kolmogorov,aronson78}.
In Fig.~\ref{fig2} we compare the
result of the MSH against the numerical results.  We see that for small
$Q_\infty$ there is a close agreement between the
numerics and the MSH result.  However, as $Q_\infty$ approaches 1
we begin to see significant departures from the MSH; the MSH predicts
that the front should be stationary for $Q_\infty>1$.  We therefore
see that the {\it linear} MSH {\it fails} to accurately
predict the selected front speed for sufficiently large trapped flux.

Further understanding of the selection problem can be gleaned from 
an {\it exact} solution of the TDGL equations for 
$Q_\infty=1$, $\kappa=1/\sqrt{2}$, and $\bar{\sigma}=1/2$. 
To motivate this solution, we first notice that when $Q_{\infty}=1$, 
it appears qualitatively that $F+Q=1$. With this in mind, 
we look for solutions to (\ref{F}) and (\ref{Q}) of the form 
$F(X) = 1 - Q(X)$; by substituting this {\it Ansatz} into (\ref{F}) 
and (\ref{Q}), we find that this is {\it only} consistent when 
$\kappa = 1/\sqrt{2}$ and $\bar{\sigma} = 1/2$.  For this set of parameters
the equation for the order parameter reduces to 
\begin{equation}
F'' + {c\over 2} F' + F^2 - F^3 = 0.
\label{exact1}
\end{equation}
The exact solution of Eq.~(\ref{exact1}) can be constructed using
the reduction of order method described by 
van Saarloos (see Ref.~\cite{vansaarloos89}, Sec.~IV), with the 
result that $c=\sqrt{2}$, and the order parameter and vector potential are
(up to translations) 
\begin{equation}
F(X) = {1\over e^{X/\sqrt{2}} + 1}, \quad Q(X) = {1\over e^{-X/\sqrt{2}} +1}. 
\label{exact2}
\end{equation}
It should be emphasized that this {\it duality} between $F$ and $Q$
is distinct from the well known duality between the 
order parameter and the magnetic field which exists in the {\it equilibrium}
Ginzburg-Landau (GL) equations at $\kappa = 1/\sqrt{2}$ \cite{duality}. 
To the best of our knowledge this is the only known exact solution 
of the TDGL equations. 
  
The exact solution serves as a useful check on our numerical results;  
our numerical solution of (\ref{f}) and (\ref{q})
for this set of parameters gives $c=1.4142$, bolstering confidence 
in the numerical work.  The exact solution also illustrates the limitations
of the linear MSH, which predicts a speed of $c=0$ for these parameters. 
Finally, the exact solution can also serve as the starting
point for perturbative solutions of the TDGL equations, along the 
lines of the renormalization group method discussed in 
Ref.~\cite{chen94}.  We find \cite{dibartolo96}
\begin{equation}
c = \sqrt{2} \left[ 1 - \frac{1}{6} \epsilon_{\kappa} 
        - \frac{2}{3}\epsilon_{\sigma} - \epsilon_{Q}  + O(\epsilon^2) \right],
\label{perturb}
\end{equation}
with $\epsilon_{\kappa} = 2 - 1/\kappa^2$, $\epsilon_\sigma = \sigma - 1/2$,
and $\epsilon_Q = Q_{\infty} - 1$.  For example, when $\kappa=0.6$, 
$\sigma=0.5$, and $Q_\infty=1$, Eq.~(\ref{perturb}) yields 
$c=1.5975$, while the numerical result is $c=1.5826$. 

Although the exact solution described above is a useful touchstone
for our numerical work, we desire a more general approach to the problem of 
calculating the front speed.  
To do this we develop a perturbative solution of Eqs.~(\ref{F}) and 
(\ref{Q}) for small $c$ (large trapped flux); 
a similar treatment for curved interfaces in two dimensions has been
given in Refs.~\cite{dorsey94} and \cite{chapman95a}.
Expand the solutions in powers of $c$ (the {\it inner} 
expansion),
\begin{eqnarray}
F(X;c) & =&  F_0(X) + c F_1(X) + \ldots, \nonumber \\
Q(X;c) & =&  Q_0(X) + c Q_1(X) + \ldots\ ,
\label{inner}
\end{eqnarray}
and substitute these expansions into Eqs.~(\ref{F}) and (\ref{Q}). 
The $O(1)$ equations for $(F_0,Q_0)$ are the 
GL equations.  The important feature of the solutions
for our purposes is that the vector potential
in the normal phase ($X\rightarrow \infty$) 
has the asymptotic behavior (up to a translation) 
\begin{equation}
Q_0(X) \sim {X \over \sqrt{2}} + {\rm e.s.t.},
\label{Q0}
\end{equation}
where e.s.t.=``exponentially small terms.'' This result shows that the
magnetic field in the normal phase approaches $1/\sqrt{2}$ (which is
$H_c$ in conventional units)---a planar interface can only
be in equilibrium when the field in the normal phase is $H_c$.
Proceeding to $O(c)$, we have
\begin{eqnarray}
{1\over \kappa^{2}} F_{1} '' & - & (Q_{0})^{2} F_{1}
    - 2 Q_{0} F_{0} Q_1  \nonumber \\
& & + [1 - 3(F_{0})^{2}] F_{1} = -  F_{0}',
\label{ordere}
\end{eqnarray}
\begin{equation}
Q_{1}'' - (F_{0})^{2} Q_{1} - 2 F_{0} Q_{0} F_{1} = -  \bar{\sigma} Q_{0}'.
\label{ordere1}
\end{equation}
The asymptotic behavior of $Q_1$ can be obtained 
as follows.  Multiply Eq.~(\ref{ordere}) by $F_0'$, Eq.~(\ref{ordere1})
by $Q_0'$, add the two equations together, and integrate the result from
$-\infty$ to $X$.  Then integrate $F_1''$ and $Q_1''$ by parts twice,
and use the
fact that $(F_0',Q_0')$ are solutions to the {\it homogeneous} versions
of Eqs.~(\ref{ordere}) and (\ref{ordere1}) (the zero mode).
The final result for the
asymptotic behavior as $X\rightarrow\infty$ is
\begin{equation}
Q_1(X) \sim  - {\bar{\sigma} \over 2\sqrt{2}} X^2 - {\beta \over \sqrt{2}} X
        + C_1 + {\rm e.s.t.},
\label{Q1}
\end{equation}
where $C_1$ is an integration constant and $\beta$ is given by
\begin{equation}
\beta  = 2 \int_{-\infty}^{\infty} \left[ (F_0')^2 +
   \bar{\sigma}(Q_0')^2 - {\bar{\sigma}\over \sqrt{2}}Q_0'\right]dx.
\label{beta}
\end{equation}
The {\it kinetic coefficient} $\beta(\kappa,\bar{\sigma})$
must be determined numerically from the
solutions of the GL equations \cite{osborn94}.

By comparing $Q_0$ and $cQ_1$, we see that these two terms become
comparable when $cX = O(1)$, indicating a breakdown of the perturbative
expansion.  This suggests introducing the
{\it outer} variable $\xi = cX$, with $f(\xi) = F(X)$ and $q(\xi)= Q(X)$.
In terms of these outer variables Eqs.~(\ref{F}) and (\ref{Q}) become
\begin{equation}
{c^2\over \kappa^{2}} f'' + c^2 f' - q^2 f + f - f^3 = 0,
\label{F2}
\end{equation}
\begin{equation}
c^2 q'' + \bar{\sigma} c^2 q' - f^2 q = 0.
\label{Q2}
\end{equation}
Expanding the solutions in powers of $c$,
\begin{eqnarray}
f(\xi;c) &=& f_0(\xi) + c f_1(\xi) + \ldots, \nonumber \\
q(\xi;c) & = & {q_0(\xi) \over c} + q_1(\xi) + \ldots
\label{outer}                    
\end{eqnarray}
we find that $f_0 = 1$, $f_1=0$ in the outer superconducting region
($\xi \rightarrow - \infty$), $f_0 = f_1 = 0$ in the outer normal
region ($\xi \rightarrow \infty$), and $q_0 = q_1 = 0$ in the
outer superconducting region, with
\begin{equation}
q_0(\xi) = A_0 + B_0 e^{-\bar{\sigma} \xi},\quad
q_1(\xi) = A_1 + B_1 e^{-\bar{\sigma} \xi}
\label{outer_sols}
\end{equation}
in the outer normal regions; $A_n$ and $B_n$ are integration constants.
The inner and outer solutions must now
be matched together in an appropriate overlap region
\cite{dorsey94,chapman95a}, with the result
\begin{equation}
A_0 = - B_0 = {1\over \sqrt{2} \bar{\sigma}},
\quad
A_1 = - B_1 = - {\beta \over \sqrt{2} \bar{\sigma}}.
\label{constants}
\end{equation}
Using this expansion, we can determine the asymptotic value of the
vector potential in the normal phase as an expansion in powers of
$c$:
\begin{equation}
Q_\infty = {1\over \sqrt{2} \bar{\sigma} c }
  - {\beta \over \sqrt{2}\bar{\sigma}} + O(c).
\label{qinfty}
\end{equation}
If we think of $Q_\infty$ as the control variable, then we have for
the selected velocity
\begin{equation}
c = {1\over \sqrt{2} \bar{\sigma} Q_\infty + \beta}.
\label{smallc}
\end{equation}

As was pointed out in \cite{dorsey94,chapman95a,osborn94} the kinetic
coefficient may actually be {\it negative} for some values of the
parameters.  One might worry that when
$\sqrt{2}\bar{\sigma}Q_\infty = -\beta$, Eq.~(\ref{smallc}) predicts
an infinite velocity,  contradicting our conjecture that
$c\le 2/\kappa$.
However, if this occurs we simply have a breakdown of the perturbation
expansion, indicating the need to keep higher order terms.
In Fig.~\ref{fig2} we compare the asymptotic result, Eq.~(\ref{smallc}),
and the numerical results, for $\kappa=1$ and $\bar{\sigma}=1$;
the kinetic coefficient for these parameters is $\beta=-0.216$
\cite{osborn94}.  The agreement is excellent in the appropriate region of
large $Q_\infty$.  The agreement is equally impressive for other
values of $\kappa$ and $\bar{\sigma}$.  In particular, when 
$\kappa=1/\sqrt{2}$ and $\bar{\sigma}=1/2$, $\beta = 0$ \cite{dorsey94},
so that the asymptotic analysis predicts $c=\sqrt{2}/Q_\infty$; if we 
set $Q_\infty = 1 + \epsilon_Q$, and expand for small $\epsilon_Q$, 
then $c = \sqrt{2}[1-\epsilon_Q + O(\epsilon^2)]$, which agrees to 
lowest order with the expansion about the exact result given in 
Eq.~(\ref{perturb}).

In summary, we have studied the propagation of fronts separating
the superconducting and normal phases, which are produced after a quench 
to zero applied magnetic field.  In addition to its
possible practical importance in understanding flux expulsion in 
superconductors, this problem provides an interesting variation on 
the theme of front propagation in unstable systems.  By varying the 
amount of flux trapped in the front we can go continuously from 
a regime at small flux in which the speed is close to that predicted
by the linear marginal stability hypothesis, to a regime at high flux 
which can be treated using the method of matched asymptotic expansions. 
While we have no general results at intermediate values of the
flux, for a particular set of parameters we have discovered an exact 
solution, which serves as the starting point for a perturbative 
calculation in this regime.  We are currently expanding our study to 
include front propagation in two dimensions and 
a detailed study of the diffusive fronts which appear 
after quenches to non-zero applied fields \cite{dibartolo96}.


We would like to thank R. E. Goldstein and M. Fowler for helpful discussions.
This work was supported by NSF Grants DMR 92-23586 and 96-28926.

\end{document}